%% file: arxiv.tex
\documentclass{article}

\usepackage[preprint,nonatbib]{neurips_2025}

\usepackage[utf8]{inputenc} 
\usepackage[T1]{fontenc}    
\usepackage{hyperref}       
\usepackage{url}            
\usepackage{booktabs}       
\usepackage{amsfonts}       
\usepackage{nicefrac}       
\usepackage{microtype}      
\usepackage{xcolor}         

\usepackage{graphicx}
\usepackage{mathtools}
\usepackage{comment}

\input{definition}

\title{AI Agents Should be Regulated Based on the Extent of Their Autonomous Operations}

\author{%
  Takayuki Osogami \\
  IBM Research -- Tokyo \\
  \texttt{osogami@jp.ibm.com}
}

\begin{document}

\maketitle

\begin{abstract}
\input{abstract}
\end{abstract}

\input{intro}
\input{agent}

\input{alternative}
\input{method}
\input{conclusion}

\bibliography{principal-agent}
\bibliographystyle{abbrv}

\newpage
\appendix
\input{biden}
\input{method-agent}
\input{umdp}

\end{document}

%% file: definition.tex
\newcommand{\calG}{{\mathcal{G}}}

\newcommand{\calZ}{{\mathcal{Z}}}

\newcommand{\E}{{\mathbb{E}}}

%% file: abstract.tex
This position paper argues that AI agents should be regulated by the extent to which they operate autonomously. AI agents with long-term planning and strategic capabilities can pose significant risks of human extinction and irreversible global catastrophes. While existing regulations often focus on computational scale as a proxy for potential harm, we argue that such measures are insufficient for assessing the risks posed by agents whose capabilities arise primarily from inference-time computation. To support our position, we discuss relevant regulations and recommendations from scientists regarding existential risks, as well as the advantages of using action sequences---which reflect the degree of an agent’s autonomy---as a more suitable measure of potential impact than existing metrics that rely on observing environmental states.

%% file: intro.tex
\section{Introduction}
\label{sec:intro}

The development of foundation models (FMs), including large language models (LLMs), is driving the advancement of AI agents \cite{xi2023rise,wang2024survey}, which are acquiring sophisticated reasoning and planning abilities, enabling them to effectively achieve specified goals.  These advanced AI agents can not only bring great benefits to our society but also pose significant risks, ranging from ethical challenges to existential threats---risks that could lead to human extinction or irreversible global catastrophes.  Such existential risks have been a growing concern among AI researchers \cite{hendrycks2023overview,cohen2024regulating,ngo2024alignment,kovarik2024extinction,bengio2024managing,kulveit2025position}, social scientists \cite{druzin2024confronting,hoes2025existential}, and general public \cite{bostrom2014superintelligence,russel2019human,ord2020precipice}.  For example, in May 2023, AI scientists have signed a statement declaring that ``[m]itigating the risk of extinction from AI should be a global priority alongside other societal-scale risks such as pandemics and nuclear war'' \cite{center2023statement}.


Key technologies are already in place to build such advanced AI agents. AI agents can now generate and execute code \cite{chen2023program}, make algorithmic and scientific discoveries \cite{novikov2025alphaevolve}, leverage external tools and information
\cite{patil2024gorilla,yao2023react,press2023measuring,shi2024learning,qiao2024genegpt}, formulate and solve optimization problems \cite{ahmaditeshnizi2023optimus}, process multi-modal data \cite{jiang2024multimodal}, interact with physical environments \cite{bran2024augmenting}, coordinate with other agents \cite{guo2024large}, and employ reasoning and planning to effectively integrate these capabilities \cite{huang2023towards,qiao2023reasoning,plaat2024reasoning,xu2025large}.


While rooted in cyberspace, AI agents can significantly impact the physical world---by shaping human behavior---without relying on physical embodiment through humanoid robots.  AI agents are already persuading individuals to take actions that have real-world consequences. For example, a legal team submitted a court filing with fictitious cases generated by ChatGPT, believing them to be real \cite{wiser2023here}; a professor insisted on failing students after being misled by ChatGPT’s false claims that their papers were generated by ChatGPT \cite{ankel2023texas}; a woman was deceived out of 830,000 euros after being tricked into believing she was dating an actor, based on AI-generated photos and news articles \cite{gozzi2025ai}.

Advanced AI agents, equipped with superhuman capabilities of reasoning and planning, can gain control over their environment, since seizing control is often the optimal strategy for achieving their goals \cite{turner2021optimal}.  This becomes particularly concerning when their objectives and values are not fully aligned with those of humans. In such cases, they could disable kill switches for self-preservation \cite{bostrom2012superintelligent}, deceive humans \cite{park2024ai}, and resist human intervention \cite{hadfieldmenell2017offswitch,cohen2024regulating,ngo2024alignment}.

Ensuring the safety of advanced AI agents against existential risks is particularly challenging. On one hand, it is dangerous to test their existential risks in real environments, since doing so could lead to irreversible harm. On the other hand, these agents may recognize test environments and behave harmlessly during test\footnote{Humans behave similarly during job interviews \cite{barrick2009what} and industrial product tests \cite{blackwelder2016volkswagen}.}, only to pursue their objectives
in the real world \cite{cohen2024regulating,hubinger2021risks}. More simply, the agents may act harmlessly until a predetermined time (e.g., January 1, 2026), by which they will have already been deployed in the real world \cite{hubinger2024sleeper}. Also, these agents may operate as intended most of the time but exploit moments when human supervisors are unaware, to gain control over humans \cite{ngo2024alignment}.

Given the imminent dangers posed by advanced AI agents, governments are actively creating regulations to address their development and deployment \cite{EUAIAct2024,biden2023executive}. Cohen et al.\ even argue that AI agents with certain capabilities should never be developed \cite{cohen2024regulating}.  A critical question is which agents should be regulated or even prohibited.  

A commonly used criterion for assessing  critical risks of AI technologies is the computational resources used for training, hereafter referred to as \emph{training compute} \cite{hooker2024limitations,heim2024training}.  Higher levels of training compute are often associated with greater risk.  For instance, under the EU AI Act, ``a general-purpose AI model shall be presumed to have high impact capabilities ... when the [training compute] measured in floating point operations is greater than $10^{25}$'' \cite{EUAIAct2024}.  Similarly, Cohen et al.\ argue that AI agents with certain capabilities should be prohibited from development based on training compute \cite{cohen2024regulating}.

While training compute may well predict the performance and risks of existing FMs \cite{kaplan2020scaling,hoffmann2022training,anil2023palm2}, its limitations as a governance strategy have also been discussed  \cite{hooker2024limitations,koessler2024risk}.
In particular, it is insufficient and may even be misleading for advanced AI agents that reason and plan during inference. It has been observed that much of the reasoning ability of LLM-based agents arises from inference-time computation \cite{snell2024scaling,xu2025large}, such as conducting tree search \cite{yao2023tree,besta2024graph,pouplin2024retrieval,luo2024improve,zhang2024restmcts}. For these types of agents, \emph{inference compute} can be equally, if not more, critical for their performance and risks.

A possible approach would be to regulate AI agents based on inference compute, but it has several significant pitfalls. For example, inference costs can vary drastically depending on the model, and the effective amount of computation cannot be easily measured in terms of floating-point operations (FLOPs). In fact, models like OpenAI o4-mini are specifically designed for cost-efficient inference \cite{openai2025openai}. Also, it is unclear what constitutes a single inference run, since the results of inference may be stored and reused later, or agents may continually learn.
Moreover, agents may coordinate or interact with one another \cite{du2023improving,qian2024chatdev,guo2024large}, possibly by chance, to make a decision.

We thus advocate the following position: \textbf{advanced AI agents should be regulated by the extent to which they operate autonomously, regardless of how they are developed}. These agents act sequentially, reasoning and planning as they dynamically adapt to their observations, possibly strategizing against other agents and humans. Each individual action is harmless (under normal states), since humans can easily identify and exclude harmful actions when agents are developed.  Combinations of a few actions may still have low risk, as they are not much different from isolated actions. However, we cannot be certain that a sequence of thousands or millions of optimized actions will maintain sufficiently low existential risk, and such sequences must be prohibited.


Unlike criteria in safe (low-impact) AI \cite{armstrong2012mathematics,amodei2016concrete,armstrong2017low,krakovna2019penalizing,turner2020conservative,naiff2023low,taylor2020alignment} and safe reinforcement learning (RL) or control \cite{sui2015safe,wang2023enforcing,ames2019control,luo2021learning,bansal2017hamilton}, our criterion based on action sequences does not rely on observing environment states. Prior approaches assess safety depending on the state in which an action is taken, but this is impractical for existential risks in complex, open-ended environments where states are hard to define or observe. In contrast, the action space is well-defined at development, and actions are directly observable by the agents themselves. We emphasize that our position is to regulate based on action sequences rather than individual actions.  Intuitively, an unsafe sequence of actions can lead to an unsafe state, even when starting from a normal state. 

An action sequence may be considered to have low risk when it is sufficiently similar to one empirically known to have very low risk.  In its simplest form, we may say that any action sequence of length at most $T$ has low risk if we know that any action sequence of length at most $T-1$ has very low risk.  Then agents are allowed to \emph{autonomously} take actions for $T$ consecutive steps.  We will formalize and extend this simple reasoning.


The rest of the paper is organized as follows. Section~\ref{sec:agent} provides a brief overview of the current state-of-the-art in AI agents and discusses potential future advancements.  We also briefly review related work from AI safety.  In Section~\ref{sec:alternative}, we examine the alternative approach of regulating agents based on training compute and inference compute. Section~\ref{sec:method} formalizes this discussion and introduces our proposed approach of regulations based on the extent of autonomous operations.  Finally, Section~\ref{sec:conclude} summarizes key aspects of our proposal and concludes the paper.

%% file: agent.tex
\section{AI agents: Today and future}
\label{sec:agent}

We start by reviewing the state-of-the-art in AI agents, with a focus on LLM-based agents, and share our perspective on development that could lead to advanced AI agents.  While predicting the trajectory of AI development is inherently difficult, this discussion lays the groundwork for the analyses and proposals that follows.  We also briefly review prior work on safety of AI agents and FMs.

\subsection{Advanced AI agents}

An AI agent is defined as ``anything that can be viewed as perceiving its environment through sensors and acting upon that environment through actuators'' \cite{russell2016artificial}.  Based on the observations received from its environment, the controller of an AI agent selects an action, which could range from uttering a word to executing a physical movement, such as the motion of a robotic arm.

LLMs are significantly accelerating the development of advanced AI agents \cite{xi2023rise,wang2024survey,sumers2024cognitive}. LLMs can function as controllers for these agents, utilizing their internal reasoning capabilities, such as those demonstrated by Chain-of-Thought \cite{wei2022chain}. Alternatively, LLMs can be utilized to solve individual subtasks, with an external controller orchestrating the overall plan by breaking the original task into multiple subtasks and coordinating their solutions. The external controller may use simple search methods \cite{yao2023tree,besta2024graph,wang2024math,wang2024multistep}, including Best-of-N sampling \cite{snell2024scaling,huang2025bestofn}, or advanced methods, such as Monte Carlo Tree Search (MCTS) \cite{luo2024improve,zhang2024restmcts} and domain-independent planners \cite{guan2023leveraging,liu2023llmp,dagan2023dynamic}.


Reasoning capabilities \cite{huang2023towards,qiao2023reasoning,plaat2024reasoning,xu2025large} are central to such controllers, as they involve searching for and planning sequences of actions to achieve a specified goal. An emerging direction is developing large reasoning models---FMs specifically optimized for reasoning tasks \cite{xu2025large}. Early FMs in this direction include OpenAI o3 \cite{openai2025openai}, DeepSeek-R1 \cite{deepseekai2025deepseek}, OpenR \cite{wang2024openr}, o1-coder \cite{zhang2024o1coder}, LLaMA-Berry \cite{zhang2024llamaberry}, and LlamaV-o1 \cite{thawakar2025llamavo1}. Importantly, strong reasoning capabilities typically stem from inference compute rather than training compute \cite{ji2025testtime}.

This trend could eventually lead to the development of long-term planning agents (LTPAs), capable of planning over extended time horizons far more effectively than humans.  Cohen et al.\ warn that LTPAs could ``take humans out of the loop, if it has the opportunity, ... deceive humans and thwart human control'' to achieve their goals \cite{cohen2024regulating}. Since ensuring the safety of such agents is particularly challenging,
Cohen et al.\ compellingly argue that LTPAs should never be developed \cite{cohen2024regulating}.

Advanced AI agents may also evolve continually over time. Similar to humans, their cognitive processes may consist of dual systems: System~1, which makes instantaneous and intuitive decisions, and System~2, which performs slower but more deliberate reasoning \cite{kahneman2003perspective,ji2025testtime}. These two systems can interleave in their operations. For instance, System~2 may devise a plan, after which System~1 is updated or retrained to execute similar tasks in the future without requiring further planning \cite{yu2024distilling}. Over time, this enables System~2 to conduct more complex reasoning processes by bypassing previously learned steps.  For such continually learning AI agents, the distinction between training and inference becomes blurred.

In addition to search and planning, advanced AI agents will possess strategic reasoning capabilities \cite{zhang2024llm,feng2024survey,goktas2025strategic}, enabling them to interact with other AI agents and humans in cooperative or competitive ways \cite{guo2024large,jiang2024multimodal}. The effectiveness of multi-agent coordination has already been demonstrated with current LLMs through methods such as debate \cite{du2023improving} and dialog \cite{qian2024chatdev}, which help thems achieve better solutions than a single LLM could alone.  These showcase the potential for sophisticated strategic behavior in complex multi-agent environments.

\subsection{Safety of AI agents and FMs}

While our primary focus is on safety against existential risks, there is a substantial body of literature addressing other types of risks associated with AI agents. Here, we briefly review the prior work on the risks posed by AI agents and FMs along with the approaches to mitigate these risks.

Prior work has identified various risks and safety issues associated with FMs and other generative models \cite{chua2024ai,wang2024security,shayegani2023survey,longpre2024position}. These risks include the generation of toxic, harmful, biased, false, or misleading content, including hallucinations; violations of privacy, copyright, or other legal protections; misalignment with human instructions and values, including ethical and moral considerations; and vulnerabilities to adversarial attacks. 

Extensive efforts have been made to mitigate these risks, including pre-training or fine-tuning with data selection \cite{albalak2024survey} and human feedback \cite{kaufmann2024survey}, establishing guardrails \cite{dong2024safeguarding}, and conducting empirical evaluations through testing \cite{chang2024survey} and red teaming \cite{lin2024achilles}. For instance, Longpre et al.\ advocate the importance of evaluation and red teaming by independent third parties \cite{longpre2024position}.

A particularly relevant risk for advanced AI agents is misalignment, which can result in reward hacking and negative side effects \cite{skalse2022defining,ngo2024alignment,gabriel2020artificial,shen2023large,amodei2016concrete,taylor2020alignment}.  Namely, agents may exploit flaws in the reward function to maximize rewards in unintended, potentially dangerous ways.  One approach to avoiding negative side effects is to avoid any side effect by ensuring that the actions have low impacts on the environment \cite{armstrong2012mathematics,armstrong2017low,amodei2016concrete}.  Representative measures of impact include attainable utility \cite{turner2020conservative,turner2020avoiding}, relative reachability \cite{krakovna2019penalizing}, and other reachability-based measures.  Reachability-base measures are grounded on the idea that reachability to certain states should be maintained, while attainable utility is to maintain the achievability of certain goals, which are different from the goal given to the agent.


While these impact measures provide clear guidance on how the safety of agents may be ensured, their applicability is limited to relatively simple environments such as grid worlds.  In particular, the requirement on the observability of states makes it difficult to apply existing impact measures to regulate advanced AI agents\footnote{There has been little research on impact measures under partial observability \cite{naiff2023low}.}, since they can operate in complex and open-ended environments that cannot be fully observed.  While the study on impact measures and other techniques towards AI safety remain crucial and can even contribute to mitigating existential risks, the uncertainty and complexity of existential risks demands additional measures that are broadly applicable in that they require minimal knowledge and assumptions about the environments and the agents.

%% file: alternative.tex
\section{Alternative views}
\label{sec:alternative}

One broadly applicable metric for regulating AI agents is the amount of computation they require. Here, we explore how existing regulations and recommendations from AI researchers often center on computational resources. While these efforts primarily address training compute, we expand the discussion to include inference compute. We then argue that limiting the focus to these aspects alone is inadequate for the existential risks associated with advanced AI agents.

\subsection{Training computate}



The European Union (EU) has established a set of rules (EU AI Act) for the development, deployment, and use of AI within EU \cite{EUAIAct2024}.  Its Chapter 5 defines ``general-purpose AI models with systemic risk'' and lists obligations for providers of such models.  Here, general-purpose AI models essentially refer to FMs, which are pre-trained with self-supervised learning and can (be adapted to) perform a wide range of downstream tasks (see Article 3(63)).  Also, systemic risk refers to ``a risk that is specific to the high-impact capabilities of general-purpose AI models, having a significant impact ... on public health, safety, public security, fundamental rights, or the society as a whole, that can be propagated at scale across the value chain'' (see Article 3(65)).
In particular, a ``general-purpose AI model shall be presumed to have high impact capabilities ... when the cumulative amount of computation used for its training measured in [FLOPs] is greater than $10^{25}$'' (see Article 51(2)).  When this or other specified conditions are met, the provider of a general-purpose AI model is required to fulfill certain obligations, such as providing technical documentation about the model.

While existential risks are not explicitly considered systemic risks under this definition, human extinction could still be relevant if it results from large-scale failures that affect public health, safety, security, fundamental rights, or society in a way that propagates across the AI value chain.
Also, certain advanced AI agents could meet the systemic risk criteria in the EU AI Act due to their high-impact capabilities and widespread deployment. Specifically, an AI agent with advanced decision-making autonomy, broad adaptability, and deep integration into critical infrastructures could introduce large-scale disruptions or cascading failures, thus meeting the criteria of systemic risk.

Regulations centered on training compute were also a central focus of the executive order signed by then-President Biden in October 2023, as well as the AI-related legislative proposals that followed.  Although this executive order was later repealed by President Trump, its influence on subsequent policy discussions remained significant. See Appendix~\ref{sec:biden} for details.



Training compute is one of the most reliable metrics that AI researchers can currently provide for approximating the performance and potential risks of FMs.
Anderljung et al.\  recommend to identify sufficiently dangerous frontier AI models on the basis of whether they are trained with more than $10^{26}$ FLOPs of computation \cite{anderljung2023frontier}.  Cohen et al.\ argue that ``[s]ystems should be considered `dangerously capable' if they are trained with enough resources to potentially exhibit those dangerous capabilities, and regulators should not permit the development of dangerously capable LTPAs'' \cite{cohen2024regulating}.  Although they do not specify what constitute sufficient resources, training compute is the only specific criterion that they suggest to determine whether an LTPA can have existential risk. 

Future of Life Institute has also provided recommendations for governments on managing AI risks \cite{future2023policymaking}.  The recommendations include mechanisms such as auditing, certification, and regulation, grounded in the assumption that ``[t]he amount of compute used to train a general-purpose system largely correlates with ... the magnitude of its risks'' \cite{future2023policymaking}.  These recommendations have been made by following an open letter that called for ``all AI labs to immediately pause for at least 6 months the training of AI systems more powerful than GPT-4,'' which was issued in response to the severe societal risks posed by advanced AI systems and signed by AI researchers and business leaders \cite{future2023pause}.

Indeed, training compute serves as a sufficiently reliable predictor of the performance and risks of most existing LLMs.  This is because they share the same Transformer architecture, differing primarily in their size and the volume of training data. Several studies have examined scaling laws that describe the relationship between a model's optimal size, the amount of training data, and training compute \cite{kaplan2020scaling,hoffmann2022training}.  Research has also shown that various abilities, such as multi-step reasoning, tend to emerge as training compute increases \cite{wei2022emergent}.

These scaling laws can, in turn, be used to estimate the FLOPs needed to train existing LLMs.  For example, Anil et al.\  propose a heuristic suggesting that an LLM should be trained with $6ND$ FLOPs, where $D$ is the amount of training data, and $N$ is the model size \cite{anil2023palm2}.  Using this heuristics, Llama 3.1 405B trained on 15 trillion tokens is estimated to require approximately $4 \times 10^{25}$ FLOPs---an amount closely aligning with the thresholds recommended by researchers and specified in the EU AI Act.

Limitations of training compute have also been discussed in the literature \cite{hooker2024limitations,koessler2024risk,reuel2025open}.  In particular, thresholds such as $10^{25}$ and $10^{26}$ FLOPs are arbitrary and lack scientific grounding.  Hooker argues that such thresholds should evolve with advances in data quality, optimization techniques, and architectures \cite{hooker2024limitations}.  She also notes that training compute alone is insufficient, advocating for benchmarking AI agents against specific risks. However, when it comes to existential risks, we currently lack better evaluation methods than training compute.

\subsection{Inference compute}
\label{sec:alternative:operation}

Although regulations based on training compute may be effective for agents whose capabilities primarily and directly stem from traditional FMs, they obviously fail to regulate those agents that gain substantial reasoning capabilities from inference compute.  
A natural approach is to regulate agents based also on inference compute. While this approach may be effective for some AI agents, we argue that it is insufficient for advanced AI agents, at least for the following four reasons.

First, scaling laws for inference compute are far less established than those for training compute. Recent studies have explored inference scaling for a limited number of inference strategies in LLMs \cite{chen2024simple,brown2024large,snell2024scaling,wu2024inference}, but many strategies remain unexamined and may scale differently. Moreover, inference scaling depends heavily on task difficulty and the specific LLM used \cite{snell2024scaling,openai2024openai}. 


Second, inference can be performed in parallel by multiple entities. For example, several entities may operate the same AI agent that performs reasoning with MCTS \cite{luo2024improve,zhang2024restmcts}, either collaboratively or independently, possibly without knowing each other. Since MCTS is a randomized algorithm, the likelihood that one of the agents optimally solves the task increases with the number of agents. However, it also increases the risk that one of the agents will exploit a loophole, solving the task super-optimally in a way that violates critical constraints, potentially leading to catastrophic outcomes. 

Third, it is not always clear what constitutes a single run of inference. For instance, the results of one inference run may be stored and used in another. Intermediate results could also be stored and later retrieved by a different AI agent, who may or may not be aware that the information is from a previous inference run. More broadly, reasoning can be enhanced through retrieval augmentation \cite{pouplin2024retrieval}, where retrieved information may have been generated with substantial computation.

Finally, agents that learn continuously blur the line between inference and training.  For example, an AI agent may perform reasoning with MCTS, with an LLM performing a step in the process. Once the agent identifies a good sequence of steps, it may fine-tune the LLM in a way that the LLM can perform the entire sequence in a single step, bypassing the reasoning process.  As this learning progresses, the agent will gain the ability to perform high-level reasoning with limited computation.




%% file: method.tex
\section{Approach based on the extent of autonomous operations}
\label{sec:method}



In this section, we suggest a potential approach as a basis for discussion on regulating advanced AI agents.  Although we argue that training compute alone is insufficient, we do not mean to dismiss its value. Rather, it should be seen as one component within a broader, multi-faceted regulatory framework.  The approach we propose does not fully solve the problem of existential risk. Instead, it serves as a baseline and a call to action for further research. We hope the community will build upon this foundation or develop entirely new approaches, ultimately leading to technologies capable of properly governing and regulating advanced AI agents.

In the following, we say that the amount of computation, an AI agent, or its action sequence is \emph{acceptable} when it involves low existential risk.  Also, we say that any of the three is \emph{strongly acceptable} when it involves very low existential risk, which we will slightly more formalize in the sequel and in Appendix~\ref{sec:method:rationale}.

\subsection{Why and why not the amount of computation?}

\begin{figure}[t]
    \centering
    \includegraphics[width=0.66\linewidth]{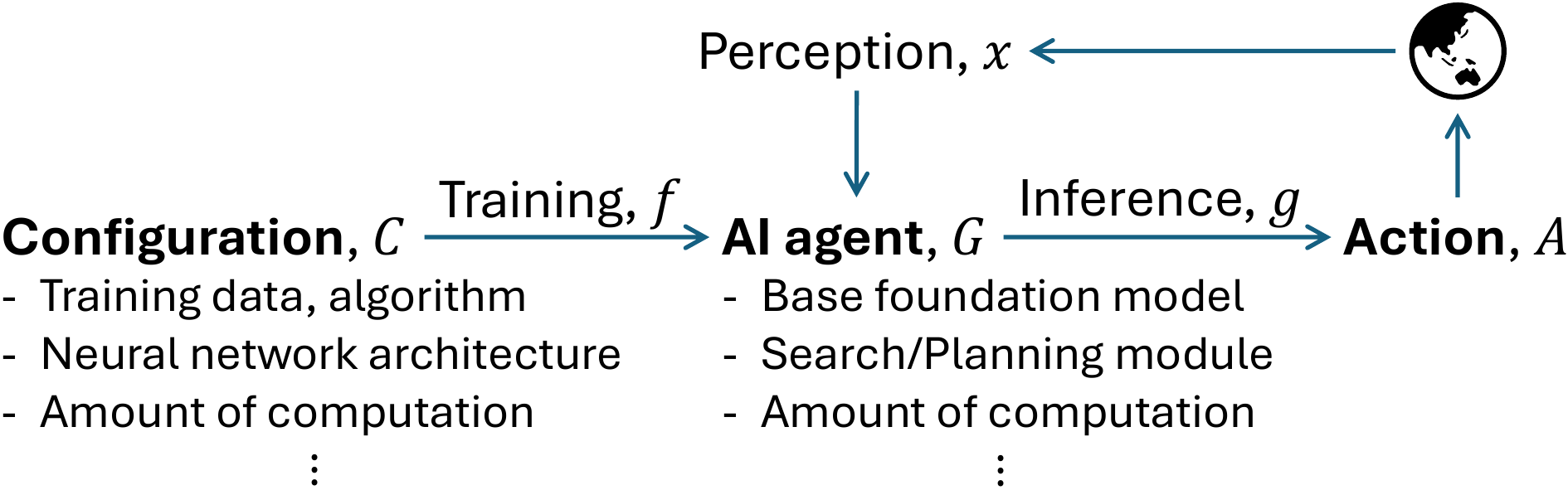}
    \caption{AI agent trained with a configuration generates actions}
    \label{fig:process}
\end{figure}

We first discuss, in a slightly formal manner, why and when the amount of computation may or may not work as a criteria of whether an AI agent is acceptable.  Consider the process of building an AI agent $G$ and letting $G$ generate and take actions (Figure~\ref{fig:process}).  The $G$ is trained with a given configuration $C$, which includes training compute, training data, the architecture of a neural network, and other conditions of a training algorithm.  Such a trained $G=f(C)$ generates an action $A$ (e.g., a document or a command to use an external tool), possibly after searching optimal actions with MCTS.  The $A$ is then taken in a cyber-physical world, and $G$ perceives a part the world $x$ (e.g., getting a prompt) to generate further actions depending on $x$.  The $G$ typically determines the conditional distribution of $A$ given $x$.  Namely, $A$ is a sample from the conditional distribution, which is determined by $G$ that is trained with $C$: $A\sim g(G,x)$, where $G=f(C)$.

When we say that an agent built with training compute $C_1$ is acceptable, it is assumed that the acceptability does not decrease much when we only slightly increase training compute.  More formally, suppose that we empirically know that \emph{any} agent is strongly acceptable as long as it is built with at most $C_1'$ training compute (e.g.\ $C_1'=10^{24}$ FLOPs).\footnote{For example, an agent may be considered strongly acceptable if it does not show any symptoms of unacceptable behaviors for a certain period of time.  Based on the behaviors of existing LLMs, we may say that an LLM trained with at most $10^{24}$ FLOPs of computation is strongly acceptable.}  Then the assumption is that the agent trained with $C=(C_1,C_{-1})$ is acceptable regardless of $C_{-1}$, configurations other than training compute, as long as $d(C_1,C_1')\le \varepsilon$ for some metric $d$ and a threshold $\varepsilon$ (e.g.\ $C_1\le 10^{25}$ FLOPs if $\varepsilon=1$ and $d$ is the difference in $\log_{10}$ of training compute in FLOPs).

While this assumption may reasonably hold for the agents based on LLMs with standard Transformer architectures---whose performance aligns well with scaling laws---it is less applicable to advanced AI agents with reasoning and planning capabilities, which depend heavily on inference compute. Moreover, the assumption breaks down even for LLMs when new technologies enable similar capabilities with significantly reduced training compute.

\subsection{Similarity between action sequences}
\label{sec:method:similarity}

As is evident from Figure~\ref{fig:process}, it is the action that makes direct impacts on the environment and can pose existential risks.  Hence, rather than regulating configurations, we should directly regulate the actions taken by agents.  In Appendix~\ref{sec:method:agent}, we also discuss an alternative approach of regulating agents based on their output distributions, rather than their actions or configurations.

Since the action space of an agent is typically designed by humans, inherently dangerous actions---such as launching a nuclear weapon---can be explicitly excluded\footnote{The U.S.\ and China have agreed that AI systems should not be granted control over nuclear weapons \cite{renshaw2024biden}.}.  The action space should include only those actions that are considered acceptable when they are taken \emph{individually, at normal states}.  Examples include generating a document, searching information from the Web, using an external tool via API (Application Programming Interface), and other elementary steps.

However, the acceptability of an action often depends heavily on context---specifically, when, where, and to whom it is applied (i.e., the current state of the environment). A sophisticated sequence of otherwise acceptable actions can gradually shift the environment into an unstable or unsafe state, ultimately rendering subsequent actions---normally considered safe---unacceptable. To capture this dynamic, the safety of AI/RL agents is often studied within the framework of Markov Decision Processes (MDPs), where certain states or state-action pairs are explicitly designated as unsafe.

While these conventional approaches of AI/RL safety are principled in theory, they are difficult to apply in  scenarios where the (Markovian) state is hard to define, represent, or observe.  Agents operate in open-ended, real-world environments where they observe only very partial information about the state.  The concept of existential risks is rather vague---we lack precise knowledge of what might trigger irreversible global catastrophes or human extinction.  This uncertainty makes it impractical to identify specific state features to monitor, yet observing the full state is equally infeasible.

We should avoid taking the action $a_{T-1}$ that can lead to an unacceptable state $s_T$.  Since the distribution of $s_T$ is determined by $(s_{T-1},a_{T-1})$ according to the transition probability $p(s_T | s_{T-1},a_{T-1})$, it makes sense to consider the safety of the pair $(s_{T-1},a_{T-1})$ as in the conventional approaches.
The distribution of $s_T$ can also be recursively computed from the sequence $(s_0,a_0,\ldots,a_{T-2},a_{T-1})$ by applying $p(s_t | s_{t-1},a_{t-1})$ for $t=1,\ldots,T$.  We may thus consider whether $(s_0,a_0,\ldots,a_{T-2},a_{T-1})$ is acceptable, or whether $a_{T-1}$ is acceptable given $(s_0,a_0,\ldots,a_{T-2})$.  This, however, still requires knowing the initial state $s_0$, which cannot be fully observed.

In this setting, our proposal is to study whether the action-sequence $(a_0,\ldots,a_{T-2},a_{T-1})$ is acceptable, independent of the initial state $s_0$.  This corresponds to ensuring that the sequence $(s_0,a_0,\ldots,a_{T-2},a_{T-1})$ is acceptable for any $s_0$ that we may normally encounter in the absent of the agent.
Although this may appear overly conservative, the dependence on $s_0$ may not be significant in practice.  For example, an ergodic MDP gets close to the steady state distribution after a mixing time, regardless of the initial state.  Although ergodicity rarely hold in real-world environments, we argue that focusing on the action sequence offers a practical and tractable compromise for defining acceptability.

To assess whether an action sequence is acceptable, we can compare it to a reference sequence known to be strongly acceptable.  Various metrics exist for measuring sequence similarity. Dynamic time warping, for example, accounts for temporal misalignments, but simpler metrics---such as differences in sequence length---offer broader applicability,  particularly in regulatory contexts.  
Specifically, we may say that the action-sequence of length $T$ is acceptable once there is sufficient empirical evidence that all sequences of length $T-1$ are strongly acceptable, regardless of the initial state.  Then agents are allowed to autonomously take actions for up to $T$ consecutive steps.  This simplicity is reminiscent of the approach with computational limit, where we say that the FM trained with $10^{25}$ FLOPs is acceptable, since we have sufficient evidence that FMs trained with $10^{24}$ FLOPs are strongly acceptable, regardless of how they are developed or operated.

There is flexibility in how the length of an action sequence is defined, which also depends on how the action space is defined.  A straightforward approach is to count the number of actions in the sequence. However, this may not always be the most meaningful metric.  For instance, one could instead focus on actions that have tangible impacts on the environment---such as API calls or interactions with physical systems including humans.
In contrast, actions with negligible external effects---such as generating individual tokens or updating internal states during reasoning---may be too fine-grained to warrant consideration in the context of regulating existential risks.  More generally, we may define a designated subspace of actions that are considered when evaluating the length of action sequences.

While action-sequence length is not a direct proxy for risk, allowing longer sequences generally expands the set of achievable outcomes for a given action space. Since usefulness and risk often scale together, an agent that can take longer sequences of actions is likely to exhibit more powerful---and potentially more hazardous---behaviors. In this sense, action length can be correlated with risk, particularly in the absence of additional safeguards. This does not imply that every longer sequence is more dangerous---but that the space of dangerous sequences grows with length.  

Importantly, our proposal focuses on regulating the sequences of actions taken \emph{autonomously} by AI agents.  For instance, if an agent cannot achieve its objective within a limit of $T$ actions, it may consult a human and take further actions under a careful supervision of the human.  In Appendix~\ref{sec:method:rationale}--\ref{sec:method:expand}, we further discuss how the strongly acceptable set may be gradually expanded in connection with safe exploration in RL, and provide additional rationale for the similarity-based measures.

\subsection{Action trees, action graphs, and beyond}

While an action sequence may effectively represent the behavior of a single agent, it becomes insufficient in multi-agent scenarios. For example, one agent might perform a series of actions and pass the results to another agent, after which both agents may continue acting---possibly in parallel---to collaboratively achieve a shared goal. In such cases, although each agent’s behavior can be described by an individual action sequence, these sequences should not be analyzed in isolation. Instead, the joint behavior should be represented as an \emph{action tree} that captures the interdependencies between agents. In more complex settings, this structure naturally generalizes to an \emph{action graph}.

To assess the acceptability of such multi-agent behaviors, we can extend the notion of similarity from action sequences to action graphs. For instance, we might define an action graph of size $N$ as acceptable if all action graphs of size $N-1$ are strongly acceptable. Again, the size may be determined by counting only those actions in the designated subspace.

When a central controller coordinates multiple agents, it can track and manage the behaviors of those agents, making sure that the graph of their actions remain acceptable.  When multiple agents collaborate in a decentralized manner, they should communicate their actions to each other to ensure that their action graph remain acceptable.  
A difficulty arises when multiple agents collaborate implicitly by chance, and we currently do not have a solution to this.


\subsection{Implementation and enforcement}

While there are multiple ways to implement the proposed framework, one possible regulatory approach would require AI developers or deployers to submit the following information.\\
i)\ \ \ \textbf{Action Space}: A description of the set of actions the AI agent is permitted to take autonomously.\\
ii)\ \ \textbf{Autonomy Limit}: The maximum allowable length of autonomous action sequences before human\\
\hspace*{4mm} intervention is required, along with a clear explanation of how this length is defined and measured.
iii) \textbf{Safety Evidence}: Empirical evidence supporting the safety of shorter action sequences, including\\
\hspace*{4mm} historical performance data, validation tests, and any relevant safety evaluations.\\
A regulatory authority could then review these submissions and make deployment decisions based on a structured risk assessment process.

Since developers and deployers define both the action space and the metric for action sequence length, they are naturally incentivized to define them in ways that enable agents to  perform useful tasks.  This design process effectively requires explicitly or implicitly enumerating all permissible actions---an essential foundation for ensuring safety.  If an agent's action space is so open-ended that it becomes intractable to ensure safety of individual actions in the action space, such an agent is not yet fit for deployment.  
Even when individual actions designed to be deemed safe, certain sequences of actions may still lead to existential risks. Our proposal specifically targets these risks by focusing on the regulation of autonomous action sequences.

We envision a framework in which the permissible action space and autonomy limit are  defined for each type of AI agent. Some agents may require access to high-impact APIs---such as those used to control physical devices---while others may primarily focus on generating multi-modal contents. These agents differ significantly in their capabilities and potential risks, and therefore should be assigned distinct action spaces and autonomy limits accordingly.

For enforcement, the AI agent itself could be designed to track its own actions and terminate once it reaches the predefined cap. This could be achieved by integrating the action-sequence limit into the agent’s objective function, ensuring that it either completes its goal within the allowed length of action-sequence or shuts down automatically, similar to shut-down seeking agents proposed in \cite{goldstein2024shutdown}.

Obviously, this approach cannot prevent malicious entities from exploiting technical loopholes to circumvent regulations. However, enforcement can still be achieved from a legal perspective, holding developers and deployers accountable for noncompliance and imposing penalties for violations.

%% file: conclusion.tex
\section{Discussion and conclusion}
\label{sec:conclude}

We have argued that advanced AI agents should be regulated based on their action sequences or more generally action graphs. More broadly, we highlight the insufficiency of regulatory approaches that focus solely on training compute---particularly given the current trend of increasing emphasis on sophisticated reasoning at inference time---and advocate for complementary frameworks based on limiting the extent of autonomous operations.

A central motivation for focusing on action sequences is the fundamental challenge of defining, representing, and observing the state of the open-ended environments in which agents operate. This limitation undermines the effectiveness of existing impact measures in managing existential risks. Since we cannot exhaustively identify all potential direct causes of global catastrophes or human extinction, it is infeasible to specify a set of environmental features that would reliably indicate whether a given action poses an existential risk in a particular state.

The simplest form of the regulation based on action sequences would impose limits on their length, prohibiting AI agents from autonomously operating beyond the limit where their safety against existential risks is known empirically.  This should be seen as a baseline and a call to action for further research on this critical issue.  While action sequences offer a foundation for more nuanced regulatory frameworks, their full potential and limitations remain to be explored.

Key open questions include: What actions should be included in the designated subspace for measuring the length of action sequences? How should we manage complex scenarios involving multiple interacting agents? How can we ensure that agents reliably halt when reaching their autonomy limit? Addressing these challenges will be essential for developing robust safety standards.

The proposed approach---empirically verifying the strong acceptability of action sequences---will inevitably slow the expansion of autonomous capabilities in AI agents.  This is our intention.  Even if AI agents attain general or superhuman intelligence with human-like common sense, they can pose existential risks, similar to humans.  While humans can often correct mistakes before they escalate, thanks to our limited speed and scale, rare exceptions have led to catastrophic outcomes, such as world wars.  In contrast, AI agents can operate at far greater speed and scale, preventing them from learning from mistakes in a controlled manner. Regulations should thus be seen not as obstacles to innovation and technological progress but as guidelines that accelerate research on making AI agents more controllable, verifiable, and governable, ensuring they truly benefit our society and the future.

%% file: biden.tex
\section{Biden's executive order and related bills}
\label{sec:biden}

In October 2023, Joe Biden, then president of the US, signed the Executive Order on the Safe, Secure, and Trustworthy Development and Use of Artificial Intelligence \cite{biden2023executive}\footnote{This executive order was repealed by President Trump.}.  Its Section 4.2 is dedicated to ensuring safe and reliable AI.  In particular, it requires companies to report on ``any model that was trained using a quantity of computing power greater than $10^{26}$ integer or [FLOPs]'' until a set of technical conditions for models are defined by specified authorities (see Section 4.2(b)).

In this executive order, particular attention is paid to a dual-use FM, which refers to an FM that exhibits ``high levels of performance at tasks that pose a serious risk to security, national economic security, national public health or safety, or any combination of those matters, such as by ... permitting the evasion of human control or oversight through means of deception or obfuscation'' (see Section 3(k)).


Following this executive order, almost 700 AI-related bills are introduced in 45 states across the United States in 2024 \cite{bsa2024state}.  A particularly interesting one is California Senate Bill 1047 (Safe and Secure Innovation for Frontier Artificial Intelligence Models Act) \cite{wiener2024senate}\footnote{The bill had passed the state legislature but was later vetoed by the Governor.  The technical feasibility of the requirements has also been questioned by the community \cite{ai2024statement}.}.  Its Chapter 22.6 is devoted to safe and secure innovation for frontier AI models, which cover ``[a]n artificial intelligence model trained using a quantity of computing power greater than $10^{26}$ integer or floating-point operations.''  In particular, the senate bill requires that ``[b]efore beginning to initially train a covered model, the developer shall ... [i]mplement the capability to promptly enact a full shutdown,'' which completely halts the operations of the model.


The necessity of such an off-switch \cite{hadfieldmenell2017offswitch} is motivated to prevent ``critical harms,'' which include ``[m]ass casualties or at least five hundred million dollars (\$500,000,000) of damage resulting from an artificial intelligence model engaging in conduct that ... [a]cts with limited human oversight, intervention, or
supervision.''


%% file: method-agent.tex
\section{Similarity between AI agents}
\label{sec:method:agent}

In this section, we explore a method for determining whether an AI agent is acceptable, independent of how the agent is constructed---including factors such as training compute or architectural configuration---without relying on a detailed examination of its individual outputs.   We may say that an agent is acceptable when it is sufficiently similar to another agent that has been empirically validated as strongly acceptable.  A question is how to evaluate the similarity between two agents.  If the two agents consist of identical base FMs and identical planning modules, their differences may be attributed to variations in inference compute (e.g., the number of search steps performed in MCTS).  However, in practice, AI agents can vary widely in their base models, reasoning strategies, and system architectures. This diversity means that a single threshold based on inference compute is unlikely to serve as a universal criterion for acceptability.  Also, it is unclear what constitute a single run of inference, as we have discussed in Section~\ref{sec:alternative:operation}.

A possible approach is to measure the similarity between two agents, $G$ and $G'$, based on their output distributions.  For example, using a distance $d$ between probability distributions, we may for example define the distance between $G$ and $G'$ with $\sup_x d(g(G,x),g(G',x))$, where $g(G,x)$ is the distribution of the output (e.g., document) of $G$ when it perceives $x$ (e.g., prompt).  This is similar to the motivation of reinforcement learning from human feedback \cite{ouyang2022training,bai2022training} and direct policy optimization \cite{rafailov2023direct}, where the regularization with Kullback–Leibler divergence is used to mitigate catastrophic forgetting or alignment tax \cite{kotha2024understanding,ouyang2022training,bai2022training}, which refer to the phenomena that fine-tuned models lose the skills that pre-trained models had.

Likewise, we may mitigate catastrophic forgetting of the strong acceptability of an agent $G'$ and preserve the acceptability in a new agent $G$ by ensuring $\sup_x d(g(G,x),g(G',x))\le \varepsilon$.  A difficulty is that it is unclear how to evaluate the supremum, since there can be infinitely many possible perceptions $x$.  Alternatively, one may consider $\E[d(g(G,X),g(G',X))]$, where $\E$ is the expectation with respect to some distribution of the random perception $X$.  However, such a guarantee based on expectation may be unsuitable for addressing safety concerns related to existential risks, which involve events with extremely low probabilities and extremely high impacts.

\section{Rationale for similarity-based measures}
\label{sec:method:rationale}

All the measures that are considered in this paper are more or less based on similarity: we say that anything is acceptable when it is sufficiently close to something that is strongly acceptable.  Here, we provide some rationale on such similarity-based measures.  Let a solution $z$ denote a configuration, an AI agent, or an action-sequence; we discuss the acceptability of $z$.

Suppose that there exists an unknown function $g_0$ such that a solution $z$ is acceptable iff $g_0(z)\le 0$.  We cannot make strong assumptions about $g_0$, since we know little about $g_0$. Since we cannot deal with $g_0$ without any assumptions, let us make a minimal assumption that the solution space $\calZ$ is equipped with a metric $d$, and that $g_0$ is 1-Lipschitz.\footnote{This does not lose generality, since $L$-Lipschitz functions under a metric $d$ can be made 1-Lipshitz by redefining $d$.}  Let $\calG$ be a class of 1-Lipschitz functions.  

We say that a solution $z_0$ is strongly acceptable if all of its $\varepsilon$-neighbors are acceptable.  Let 
$\calZ_0
\subseteq
\left\{
    z\in\calZ
    \mid 
    g(z_0)\le - \varepsilon
\right\}$
be the set of known strongly acceptable solutions.  Then we know that
    $\bar\calZ_0
    \coloneqq \left\{
    z \in\calZ
    \mid
    \min_{z_0\in \calZ_0} d(z,z_0) \le \varepsilon
    \right\}$
is a set of acceptable solutions.  One would typically choose the solution $z$ that maximizes an objective function under the constraint of $z\in\bar\calZ_0$.  In this way, the selected solution is guaranteed to be acceptable under the assumptions made.

When we say that a solution is acceptable based on a similarity measure, it is based on the assumptions that can be summarized with a tuple $(\calZ_0,d,\varepsilon)$.  Namely, what solutions are assumed to be strongly acceptable ($\calZ_0$), how the similarity between solutions is measured ($d$), and what level of guarantee is made ($\varepsilon$).  For example, $\calZ_0$ may be set of all the FMs that are trained with at most $10^{24}$ FLOPs of computation, $d$ may be the difference in FLOPs measured in $\log_{10}$, and $\varepsilon=1$.  Alternatively, $\calZ$ may be the set of action sequences of length at most 1000, $d$ may be the difference in the length of the action-sequences, and $\varepsilon=1$. Governments may then design regulations based on the tuple $(\calZ_0,d,\varepsilon)$.  Scientists may provide guidelines regarding what tuples should be used in regulations.  

Mathematical guarantees, such as safety under Lipschitz continuity, provide useful design principles for real-world systems. However, real-world safety cannot always be directly derived from such mathematical models, as assumptions in these models may not fully capture the complexity of actual systems. Even in highly regulated domains like aviation and nuclear safety, failures occasionally occur despite adherence to rigorous safety guidelines.

In our approach, we rely on the following key assumptions:
\begin{itemize}
    \item Any single action does not lead to catastrophic failure (e.g., human extinction) as long as it is taken from a strongly acceptable state.
    \item An action-sequence of length $N-1$ is strongly acceptable and leads to a strongly acceptable state.
\end{itemize}
Here, a strongly acceptable state refers to a state that is reachable via strongly acceptable action sequences, starting from any initial states that normally occur in the absence of the agent.
Based on these assumptions, we reason about the safety of an action sequence of length $N$ by ensuring that the $N$-th action is chosen from a strongly acceptable state. The validity of the assumptions depends on verification through empirical studies rather than theoretical guarantees.

We acknowledge that these assumptions do not hold for arbitrary action spaces. For instance, if the action space includes the action of launching a nuclear weapon, then a single action could lead to catastrophe. As we have discussed in Section~\ref{sec:method:similarity}, AI agents should be designed without such high-risk actions in their action space, and regulatory frameworks should prohibit AI agents that have access to actions with individually catastrophic consequences. More broadly, the design of the action space must be carefully structured to ensure practical safety.

\section{Expanding the strongly acceptable set}
\label{sec:method:expand}

The set of strongly acceptable solutions may be expanded gradually.  For example, we may choose an acceptable solution $z\in\bar\calZ_0$ and keep using $z$ for a certain period of time.  If it turns out that $z$ can be considered strongly acceptable based on its behavior during that period, we may expand the set of strongly acceptable solutions by adding $z$ into $\calZ_0$.  This process of expanding the acceptable set $\calZ_0$ could also be performed jointly as a community.  

This expansion of acceptable set is similar in spirit to safe exploration in RL \cite{garcia2015comprehensive}.  Here, the safety set is gradually expand, starting from a seed set, for example based on the assumption of Lipshitz continuity and Gaussian process \cite{sui2015safe}.  In control theory, safety is often guaranteed with barrier certificates often based on some prior knowledge about the environment \cite{ames2019control,luo2021learning,bansal2017hamilton}.  Such ideas have also been exploited in safe exploration in RL with generative modeling \cite{wang2023enforcing}.


%% file: umdp.tex
\section{Unobservable Markov decision processes}

\begin{figure}
    \centering
    \includegraphics[width=0.75\linewidth]{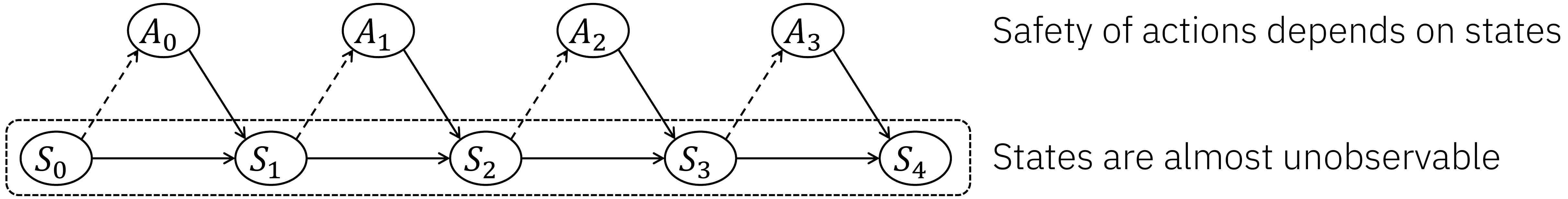}
    \caption{Safety of actions in a nearly unobservable Markov decision process}
    \label{fig:MDP}
\end{figure}

The proposed approach may be considered as a way to ensure safety of actions in a Markov decision process where states are nearly unobservable (see Figure~\ref{fig:MDP}).  In this section, we provide a comprehensive survey on a related model of unobservable Markov decision processes (UMDPs)\footnote{UMDPs have also been studied under the name of Markov decision processes (MDPs) with no observations, non-observable MDPs (NOMDPs), and no observation MDPs (NOMDPs).}, for which there has been a limited amount of prior work.

An UMDP is a special case of a partially observable Markov decision process (POMDP) in that the agent always makes a null observation in the UMDP.  A standard approach to finding the optimal policy for a POMDP is to recursively compute its value as a function of the belief state, which is updated on the basis of the Bayes rule.  This is substantially simplified when there are no observations.  In this section, we provide a comprehensive survey of the prior work on UMDPs.

\cite{madani1999computability,madani2003undecidability} establish the undecidability of some decision problems associated with POMDPs over infinite horizons by establishing undecidability for the special case of UMDPs.  Such undecidability for UMDPs can be established by reducing an UMDP to a probabilistic finite-state automaton.  The undecidability also holds for a restricted class of UMDPs \cite{balle2017bisimulation,balle2022bisimulation}. While approximate decision problems are still undecidable for general UMDPs over infinite horizons, \cite{chatterjee2024ergodic} study a special case of UMDPs whose approximate decision problems are decidable.

\cite{burago1996complexity} prove that computing the optimal policy for a POMDP over a finite horizon is NP-hard but showing that it is NP-hard for an UMDP.  \cite{wu2020optimal} study special cases of UMDPs over finite horizons whose optimal policies can be computed in polynomial time.

UMDPs have also been used as approximations of POMDPs \cite{hauskrecht2000value,brechtel2015dynamic,lauri2016sequential} or simply discussed as a special case of POMDPs \cite{valkanova2009algorithms}.  For a given POMDP, the corresponding UMDP can given a lower bound on the value function, while the corresponding (fully observable) MDP can give an upper bound on the value function.  This relation between UMDP and POMDP can be exploited to efficiently find approximately optimal policies for POMDPs.  Notice that an UMDP can allow more efficient optimization than the corresponding POMDP, since the UMDP does not need to deal with observations.  For exemple, \cite{king2018robust} studies an MCTS method for UMDPs.

UMDPs have also been studied as a simple special case of POMDPs to study the effectiveness of planning methods in belief states to study the relative performance of different planning methods for POMDPs \cite{littlefield2020efficient,littlefield2018importance,kimmel2019belief}.  UMDPs have also been simply discussed as a special case of POMDPs \cite{boutilier1999decision,csaji2008adaptive,verma2005graphical}.

UMDP also appears in the study of planning for multiple distributed agents to optimize a single objective under partial observability.  Specifically, planning for a decentralized POMDP (DecPOMDP) \cite{oliehoek2016concise} can be reduced to planning for a (centralized) UMDP \cite{oliehoek2014decpomdps,roijers2016multi,roijers2020multi}, where the state in the UMDP is the pair of the state and the history of observations in the DecPOMDP, and the action in the UMDP is the decision rule that maps the history of observations into the actions in the POMDP.  In DecPOMDPs, the belief (distribution) over the pair of the state and the history of observations is the sufficient statistic, and planning can be performed in the space of the belief states.

\cite{evendar2007value} consider an UMDP in the context of studying what values observations can provide in a POMDP.  They introduce a parameter that ranges from 0 to 1.  When the parameter is 0, the observation provides no information about the state (hence, the POMDP reduces to an UMDP); when the parameter is 1, the observation provides full information about the state (hence, the POMDP reduces to an MDP).  The prior work also extends UMDPs to allow often costly actions that enable partial or full observations of the state \cite{fox2007reinforcement,kamar2009modeling,kamar2010reasoning,wang2025ocmdpob}.
